# Glow Discharge AES: Methodological Peculiarities of Pulse Element Analysis and Flash Desorption

Vasil G. Bregadze∗, Eteri S. Gelagutashvili, Ketevan J. Tsakadze.

Andronikashvili Institute of Physics, 6 Tamarashvili st., Tbilisi, Georgia 0177;

∗Address for correspondence:
 e-mail:  **breg@iphac.ge**
           **v_breg@yahoo.com**

**Abstract**

Different techniques of Glow Discharge AES are described in this paper. The most important parameters at such investigations are: the power of VHF-field, pressure of the inert gas and concentration of the easily ionizable additive, e.g. NaCl. The influences of these parameters were studied

It is proposed a high sensitive flash desorption method, which enables investigation of the water desorption rate from humidified samples of biological origin, namely DNA and chromatin.

The ways of minimizing of detection limit are considered as the most important characteristics of an analytical device. The concentration of any measured element is detectable if it correlates to the signal equal to tripled standard deviation of the results of background measurement.

Electron temperature of the Helium has been evaluated by absorption rate at two lines of helium $\lambda = 353.828$ nm; $\lambda' = 344.759$ nm that was equal to $T_e \approx 15000$ K.

# Introduction

The basic idea of the AES method can be illustrated as follows:

A + energy $\to$ A* $\to$ A + $h\nu$, (1),

where A and A* are the atoms in the gas phase in the ground and excited electron states respectively; $h\nu$ is the quantum of emitted light at A* $\to$ A transition. The atom excitation in this case is caused by collision between electrons, atoms, and molecules of the sufficient kinetic energy. Thus, to make scheme (1) work, it is necessary to atomize the analyzed substance and provoke the subsequent atom excitation.

Generally, the source of atom excitation in AES is the source of atomization at the same time. Electrodeless discharge can be of either capacitive or inductive character. The latter is used more widely. Elcrtordless plasma of both atmospheric and reduced pressure (1-20 Torrs) is used in AES. Below we use the term electrodeless plasma to imply inductively coupled plasma (ICP) induced by electromagnetic fields of RF, VHF and microwave range.

Under conditions of atmospheric pressure, the ICP state is close to thermodynamic equilibrium. In this case, the fraction of excited atoms can be determined from the Boltzman distribution:

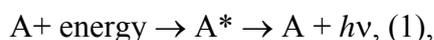

$$n_1/n_0 = K \exp(-\Delta E/kT), \qquad (2)$$

where $n_1$ and $n_0$ are the numbers of atoms in excited and ground states respectively; K is a coefficient, depended on statistic weights of upper and lower states for o → 1 transition of the given atom; ΔE – energy difference between the ground and excited states equal to ΔE = $h\nu$; k – the Boltzman constant; T – absolute temperature.

Distribution (2) shows that even at T ~$10^4$K (kT ≅ 1eV, little number of atoms is excited. The plasma temperature of 10 000 – 11000°K can be obtained by a plasmatron using RF [1-3 (20-22)] discharge. Generally, in order to obtain the mentioned temperatures, a RF generator of 1 – 30 MHz frequency and 1 – 3 kW power [4 (23)] are used. In this case, 20 liters per minute of gaseous argon is required [4 (23)].

**Experimental Part**

**Pulse VHF Inductively Coupled Plasma Spectroscopy for Element Analysis in the Solutions.** As it has been mentioned, VHF ICP has not been used earlier as a source of light for optical spectral analysis despite its obvious advantages (stability, good integration with the units of the optical systems). In the Institute of Physics the atomic-emission spectrophotometers were designed on the basis of VHF plasma of reduced pressure allowing to carry out elemental analysis (of the human peripheral blood as well) in 3-5 $\mu l$ of the solution with the detection limit of $10^{-10} - 10^{-12}$ g in absolute values.

**Block-diagrams of the monochannel spectrometer**. Figure 1 shows the block-diagram of the pulse spectrometer.

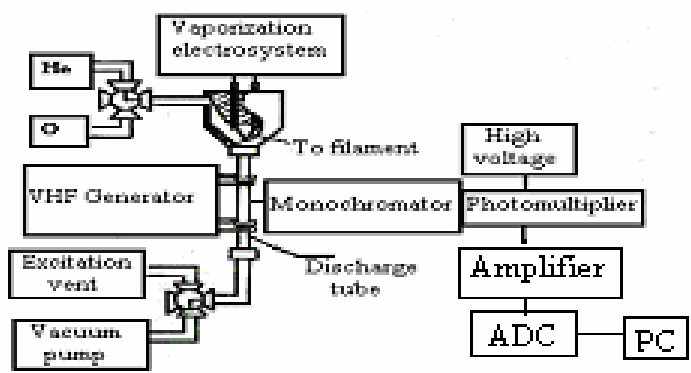

Fig.1 block-diagram of the pulse spectrometer.

$3\mu l$ of an analyzed solution is placed into the vaporizer where it is dried, ashed and vaporized with the help of the gas system and Joule heating and then injected into the discharge tube in the aerosol state. There the investigated sample of inorganic matter is atomized and excited. Optical system adjusted to the measured elements registers the light flows. The quantitative analysis is carried out by the comparison of light flows emitted by the sample and the standard. First, let us consider the common units of these spectrometers – gas-vacuum chamber, vaporization system and vaporization electrosystem.

**Gas-vacuum system.** It is designed to provide stable, thoroughly controlled, and easily obtained flows of helium and air for drying and dry ashing of the organic matter of the analyzed sample. It also creates necessary optimum pressure and provides the blow-out of the aerosol through the quartz tube of the plasmatron.

It should be specially noted that the presence of the oxygen or air lines is necessary for the element analysis of various biological samples, especially of the human blood serum [5,6 (37, 41)]. The absence of oxygen leads to the formation of incomplete combustion products (soot), which leads to the spread of he results.

**The vaporization chamber** is a stainless steel cylindrical reservoir with a discharge quartz tube geometrically inserted in its bottom. The chamber is closed by Teflon (PTFE) plug with two incorporated copper tubes-electrodes. The V-shaped tantalum filament of the length ~20 mm, diameter – 0.2 mm and 1mm radius is fastened to the end of the tubes-electrodes with the help of collect clips. The electric resistance of the filament is 0.3 ohm at the room temperature. The union connecting the cylindrical vaporization chamber with the gas system is soldered in perpendicular to the chamber. The vaporization chamber with the discharge tube piercing the inductive circuit is adjusted to the optical rail with the help of the stand with the rider.

**The vaporization electrosystem** is intended for the Joule heating of the tantalum filament for drying, dry ashing and vaporization of the analyzed sample and also for the complete vaporization of the residues after the measurements. Discharge the capacitor bank of the total capacity of 0.2F is caused by commutation of the discharge circuit with diode thyristor.

**Optical System and the system of photosignal recording**. In the monochannel spectrometer the emission of the analyzed element is focused by the two-lens quartz condenser to the entrance slit of the fast monochromator MDR-2 (Russia) with the aperture ratio 1:2.5. During the analysis of Mg, Ca, Cr, Fe, Co, Ni, Cu, Zn, Cd, Pb in the monochromator the diffraction grating (replica) with 1200 lines per 1 mm was used. In this case, the reciprocal linear dispersion MDR-2 was 2 nm/mm. The entrance and exit slits of the monochromator were generally equal to 0.05 mm. The image of the exit slit was focused by two-lens condenser to the photocathode of the photomultiplier. Unit B5-24A (Russia) served as a high-voltage power source. The signal from the photomultiplier was recorded by universal storage oscillograph S8-13 (Russia). To tune the spectrometer to the necessary line, we used the high-frequency spectral electrodless lamp, in which the discharge was excited with the help of VHF generator PPBL-3 (Russia), which was placed into the resonator instead the plasmatron discharge tube. Undoubtedly, this is one of the advantages of our device in comparison with analogous devices with the microwave plasmatrones of reduced pressure, where it is necessary to disassemble the microwave plasmatron in order to tune it. In addition, pulse VHF-plasma spectrometer was connected with the measurement-computer system (IVK-1-Russia) based on the computer SM-3 (Russia) through analog-digital converter (ADC). It is noteworthy that the measurement of the signal amplitude is more adequate to the concentration of the element under determination, than the area below its peak. Analogous conclusion was made by the authors in [7 (9)] as well.

**The technology of measurement and its control unit.** The sequence of operations of the element analysis is schematically shown in Fig.2.

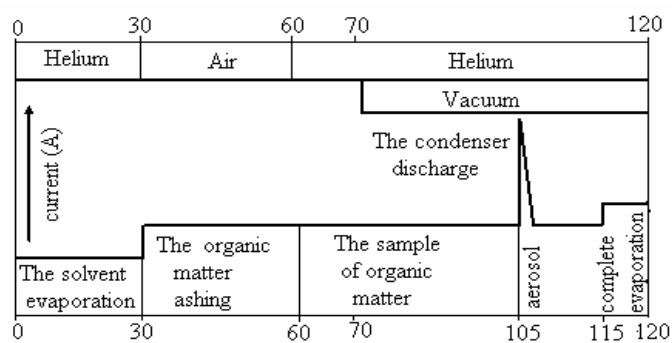

Fig. 2. The sequence of operations for element analysisby the method of pulse VHF ICP os emission spectrophotometry

The sample, e.g. 3 μl of an investigated solution is placed on the tantalum filament in a chamber where homogenous low-pressure flow of helium is established. The time count starts simultaneously with the plugging in the capacitor bank and, accordingly, heating of the tantalum filament at I = 1.5 – 2 A). During 20- 30 sec. the sample dries up. Then the chamber is connected by the valve to the air line for dry ashing. Then the air line for dry ashing of the organic matter is connected with the help of gas valve connector. At the same time, increase of the current through the filament up to 2-3A is necessary. The ashing time is equal to 20-40 sec. followed by disconnection of he air line and connection of the helium and the vacuum lines. After 5 seconds, VHF field of a generator and Tesla coil for ignition of plasma is switched on. In 5-10 sec., plasma in the quartz tube becomes stable and the spectrometer is ready for measurements. After 100 sec of starting of technological procedures, the discharge capacitor bank is to be disconnected from the power source. On the 105th second the thruster connects the capacitor bank to the tantalum filament, which is alive and simultaneously with the battery discharge, the signal rate and shape is registered (Fig. 3).

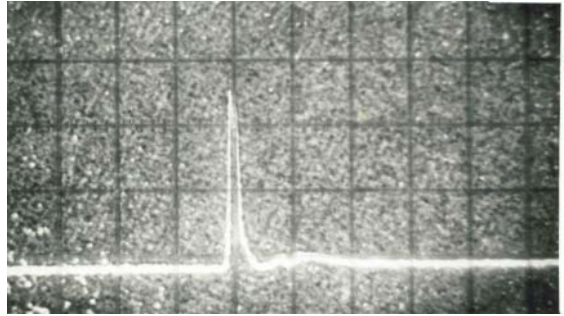

Fig. 3. Oscillogram obtained at emission analysis of Cu in the $CuCl_2$ solutions by the pulse method.
$C_{Cu}$=1.7µg/Ml, $C_{NaCl}$ = 0.15M, $\lambda$ = 324.75nm.
Amplification – 50mV/div.  Time sweep – 100 msec / div.

Then the current should be increased (fig. 2) up to 3.5-3.8A in order to increase the tantalum temperature for complete vaporization of the organic matter, which takes 5 seconds. Then VHP field, vaporization electrosystem and vacuum line are disconnected and the device is ready for the next analysis.

**The block-diagram of the spectrometer for flash desorption**. Gas-Vacuum system and VHF plasmatron are similar to the described above. The only difference is that teflon ampoules with the inner volume of 6-1000µl (depending on the mass of the sample investigation) serve as a vaporization chamber (Fig. 4).

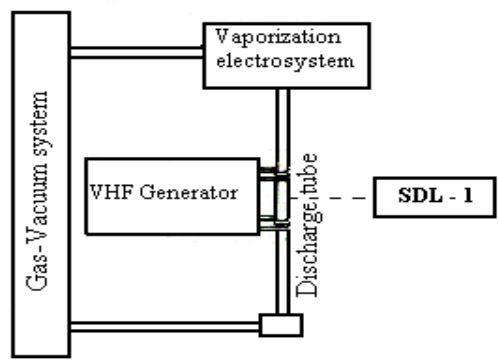

Fig.4. AES block-diagram for kinetic investigations with VHF-plasmatron of reduced pressure.

The ampoule is inserted into the vessel flowed over by the thermostatical liquids to maintain the constant temperature in the experiments on water desorption.

At the pressure 1 Torr, the discharge in the quartz tube with flowing helium is focused by the condenser to the entrance slit of the double diffraction monochromator of the spectrometer SDL-1 (Russia). The outlet of the amplifier is connected with recorder and digital voltmeter. The digital voltmeter is connected to PC.

### Results and Discussion

The detection limit – the minimum detectable concentration of the analyzed element – is the most important characteristics of an analytical device. The minimum reliable detected concentrations is equal to:

$$C_{DL} = \frac{\left(X_{DL} - \bar{X}\right)}{H}, \qquad (3)$$

where $X_{DL}$ is average value of the physical parameter; $\bar{X}$ – average value of the physical parameter in the open circuit test; H – sensitivity, determined from the slope of the calibration chart $H=(\delta X/\delta C)_{Cu}$; $X_{DL}= \bar{X} +kS$, where k is the coefficient characterizing the reliability, e.g. determined according to Student's criterion and taken as equal to 3, which means confidence level 99,86% at purely Gaussian distribution, S _ standard deviation of the results of the calculating the open circuit test. Thus, $C_{DL}=3s/H$.

So, the concentration of any measured element (substance) is considered detectable if it correlates to the signal equal to tripled standard deviation of the results of background measurement.

The detection limit of element concentration depends strongly on signal-noise ratio. So, the minimum detection limit end best reproducibility is supported by maximization of signal-noise ratio. Generally, the investigator seeks for the maximization of the signal, which is most easy and accessible. The quantity of the analyzed element in the sample is proportional to the area below the photoelectric signal under the condition of linearity of the calibration chart. In our case of pulse VHF-plasma emission spectroscopy the shape of the signal represents the passing of the substance under investigation through the discharge, which reminds the desorption spectrum of flash-desorption. The oscillograph helps to record the amplitude of the photoelectronic signal more precisely and reliably than the area below it. So, in our case, the maximization of the signal means the maximization of the area below it and subsequent signal narrowing in time. The former is caused by the efficiency of atomization and excitation of the atoms of the analyzed sample by the temperature of plasma and electron density in it (pressure, power of the VHF-field, the position of the discharge tube with respect to the resonator, the nature of the inert gas and concentration of the easily ionizable alkaline additive); the latter depends on the process of ample preparation for the analysis (current intensity and the time of drying and ashing), vaporization velocity (the voltage of the capacitor bank and the resistance of the tantalum filament) and on the time of presence of aerosol in plasma (the velocity of gas flow). Besides, as the detection limit depends on the signal-noise ratio, it is possible to increase this ratio using the high-resolution monochromator to distinguish the spectral line, which allows carrying out the analysis with the sufficiently wide slits. The low-noise photomultiplier also promotes the increase of the signal-noise ratio.

Thus, it is necessary to optimize at least 14 parameters in VHF induced plasma emission spectrometry for reliable high-sensitive element microanalysis. The most important parameters for many elements studied by biology are: the power of VHF-field, pressure of the inert gas and concentration of the easily ionizable additive, e.g. NaCl. Figures 5, 6 and 7 demonstrate the influence of this parameters on the intensity of Cu line 324.75 nm.

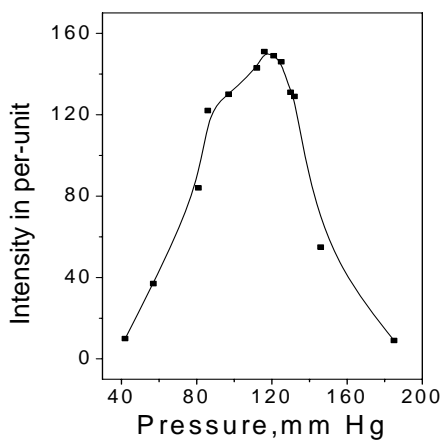 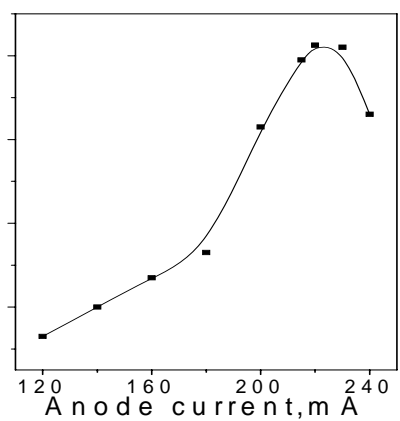

Fig. 5. The influence of the helium pressure.   Fig.6. Influence of the VHF field power.

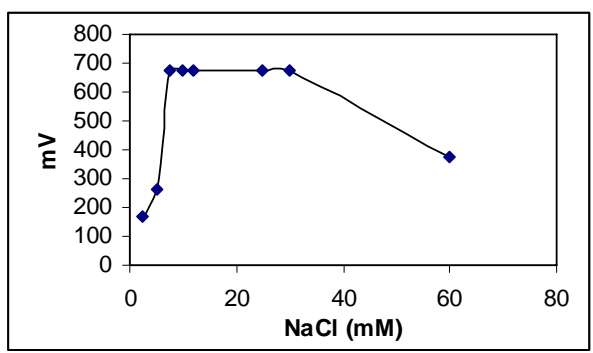

Fig.7. The influence of NaCl concentration .

The detection limits of the elements under study: Mg, Ca, Fe, Co, Ni, Cu, Zn, Cd and Pb with the spectrum lines used for their determination are given in Table 1.

Tab. 1. The detection limits of different elements

| Elements | λ, Å | Detection limits, pg |
|---|---|---|
| Mg | 5184 | 0.4 |
| Ca | 4227 | 0.2 |
| Cr | 4254 | 13.0 |
| Fe | 3735 | 5.0 |
| Co | 3454 | 18.0 |
| Ni | 3619 | 35.0 |
| Cu | 3248 | 1.0 |
| Zn | 6362 | 0.8 |
| Cd | 6439 | 2.0 |
| Pb | 4058 | 90.0 |

A certain advantage of the method comprises ability of accumulation of the analyzed substance on the tantalum filament by adding a new portion of the sample without vaporizing the previously ashed material on it. As our measurements have shown, four time addition is quite reasonable. After that, excessive accumulation of Na atoms existent in solutions of biomolecules lowers accuracy of the measurements. When number of additions is less then 4, the obtained signal linearly increases and the initial concentration can be calculated by simple division of the result on the number of additions (Fig. 8).

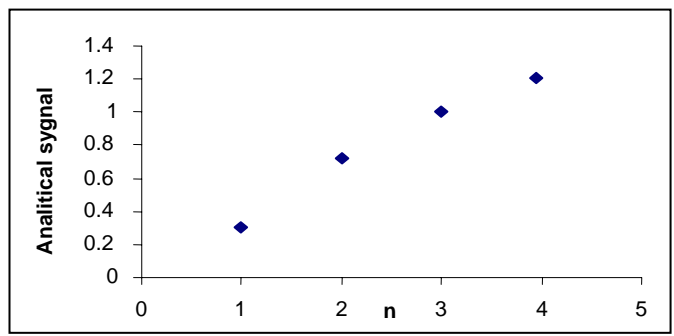

Fig. 8. Dependence of analytical signal on number of accumulations.

**Measurement of the Temperature**. If we assume the helium plasma to be homogeneous and clear, its electron temperature $T_e$ can be evaluated by means of absorption rate at two spectral lines of He using the formulae [8 (29)]:

$$T_e = \frac{E_2' - E_2}{\ln I \lambda^3 g_1' f_{12}' - \ln I' \lambda'^3 g_1 f_{12}}, \quad (4)$$

where $\lambda = 353.828$nm; $\lambda' = 344.759$nm.

Table 2. shows energies of the excited and ground levels of these lines ($E_1$ and $E_2$), their statistical weights ($g_1$ and $g_2$), and oscillation forces ($f_{12}$ and $f_{12}'$) for the transitions $E_1 \to E$;

Tab. 2. Energies of the excited and ground levels ($E_1$ and $E_2$), their statistical weights ($g_1$ and $g_2$), and oscillation forces ($f_{12}$ and $f_{12}'$) for the transitions $E_1 \to E$;

| $\lambda$, nm | $E_1$ | $E_2$ | $g_1$ | $g_2$ | $f_{12}$ |
|---|---|---|---|---|---|
| 344.759 | 20.61 | 24.21 | 1 | 3 | 12 |
| 325.828 | 20.61 | 24.42 | 1 | 3 | 3.1 |

Considering that for the total intensities $I = 46$ and $I' = 132$ in relative units, we get $T_e = 1.4$eV that corresponds to $T \approx 15000$K. Analogous calculations were carried out for the air plasma of reduced pressure using lines of Balmer series $H_\alpha = 656.284$nm and $H_\beta = 486.133$nm. The value of the corresponding electronic temperature was $T \approx 3000$K.

**Vaporization chamber in the regime of flash-desorption.** The ambiguity of the connection of the humidified sample to the spectrometer, which might cause the plasma instability, is the disadvantage of the isothermal desorption. The rapid opening of the valve may even cause plasma suppression. These disadvantages are not important and can be taken into consideration. They can be even eliminated in the flash-desorption mode (i.e. when recording the temperature spectrum of desorption), where the role of the valve is played by the sufficiently low temperature allowing significant decrease of desorption (comparable with the background). The temperature rise causes graduate increase of desorption rate, then the temperature reaches its maximum and then finally decreases to the background level. The rate is low both at the beginning and in the end (Fig. 9). Though there is much water in the beginning, both the temperature and the vapor pressure above the sample are low, while at the end the pressure is low due to the great decrease of the amount of hydrated water.

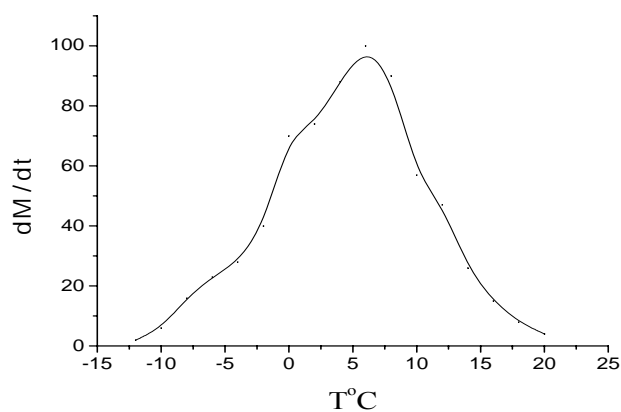

Fig.9. Spectrum of flash-desorption of water from Na DNA humidified under equilibrium conditions at 92.5% r.h. (0.84 g $H_2O$ / g of dry weight of Na DNA)

Flash-desorption is a powerful method of the study of adsorption and desorption rates at the investigation of physics and chemistry of the surface [9 (42)]. One of the variants is the mass-spectrometry of flash desorption, which is a rather complicated method from the point of view of the experiment [10,11 (43,44)]. The proposed method of desorption (including flash-desorption) has higher sensitivity, in some cases exceeding the sensitivity of mass-spectrometry and is much more simple. It is noteworthy that it is the first investigation of the water desorption rate from the humidified samples of biological origin, namely DNA and chromatin, and it is difficult to imagine any other method permitting to investigate this parameter.

The simplicity of the constant temperature maintenance is one of the advantages of isothermal desorption. The experiments at flash-desorption require the exact knowledge of temperature and its linear increase in time allowing simple calculation of the activation energy of water desorption. Fig.8 shows the spectrum of flash-desorption of the heavily humidified at 92.5% r.h. and 25ºC sample of Na-DNA (0.84g of water per gram of DNA).

## Acknowledgments

The authors are very grateful to Prof.N.Kirvalishvili and Prof. D Tskhakaia for helpful discussions on problems connected to gas discharge and Prof. Dr. Monaselidze for permanent discussions on issues of molecular biology of cell, especially, on affects on DNA leading to point mutations. We would also like to thank Prof. S. Nanobashvili for discussions on microwave discharge.

**References:**

1. 1 V. Kh. Goikman, V.M. Goldfarb. in: *Plasmo-Chemical Reactions and Process*, M. Nauka. 1977, pp. 232 – 278;
2. M.E.Britske, V.M. Borisov, Yu.C. Sukin. Zavodskaia Lab., 1967, vol. 33, pp. 252-256;.
3. Yu.P. Raiser. *Laser Spark and Propagation of the Discharge*. M. Nauka, 1974, p. 308;.
4. V.A. Fassel, *Science*. 1978, vol. 202. 4364, pp. 183-191;
5. V.G. Bregadze E.S. Gelagutashvili., the method of determination of Cu content in blood serum. Georg. pat., 128 69 89, B. I. , 46, 1986;
6. V.G. Bregadze in: *New Physical Methods in Biological Investigations*, Nauka, Moscow, 1987, 33-45;
7. E.D.A. Stemp, and J.K. Barton in *Metal Ions in Biological Systems*, eds. H.Sigel and A.Sigel, Marcel Dekker Inc., New York, Basel, 1996, vol. 33, ch. 11, pp. 325 – 365;
8. G.A.Kasabov, V.V. Eliseev. in *Spectroscopic Tables for Low Temperature Plasma*. M.: Atomizdat, 1973, p.160;
9. A.W. Adamson. in *Physikal Chemistry of surfaces*, M. Mir. 1979, 568;
10. P.W. Tamm, L.G. Schmidt, *J. Chem. Phys*. , 1969, vol. 51, pp. 5353-5363;
11. P.W. Tamm, L.G. Schmidt, *J. Chem. Phys*. 1971, vol. 54, pp. 4775-4787.